*Article*

# Application of Methods of Artificial Intelligence in Systems for Continuous Automatic Monitoring of Dust Concentration and Deposits in Mine Atmosphere

**Daria Trubicina** [1,2]**, Kirill Varnavskiy**[1]**, Alexander Ermakov** [1]**, Fedor Nepsha** [1,3]**, Roman Kozlov** [1]**, Naser Golsanami** [4] **and Sergey Zhironkin** [5,6] *****

[1] Mining Industry Digital Transformation Lab, Mining Institute, T.F. Gorbachev Kuzbass State Technical University, 28 Vesennya st., 650000 Kemerovo, Russia; varnavskijka@kuzstu.ru (K.V.); nepshafs@kuzstu.ru (F.N.); ermakovan@kuzstu.ru (A.E.)

[2] Gorniy-TSOT" Ltd, Sosoniy bulvar, 1, 650002, Kemerovo, Russia;

[3] Department of Theoretical Electrical Engineering and Electrification of Oil and Gas Industry, Gubkin University, 119991 Moscow, Russia

[4] State Key Laboratory of Mining Disaster Prevention and Control Co-founded by Shandong Province and the Ministry of Science and Technology, Shandong University of Science and Technology, Qingdao 266590, P. R. China

[5] Institute of Trade and Economy, Siberian Federal University, 79 Svobodny av., 660041 Krasnoyarsk, Russia

[6] Department of Open Pit Mining, T.F. Gorbachev Kuzbass State Technical University, 28 Vesennya st., 650000 Kemerovo, Russia

***** Correspondence: zhironkinsa@kuzstu.ru

**Abstract:** With the growth of coal production, the load on the production capacity of coal enterprises also increases, which leads to a concomitant increase in dust formation in both opencast and underground methods of mining coal deposits. Dust, generated during drilling, blasting operations, excavation, loading, crushing and transportation of mined rock is one of the factors that has a negative impact on the health of mining workers and on the level of environmental pollution with solid particles. Thus, increasing the efficiency of controlling the concentration of solid particles in the mine atmosphere and dust deposits is an urgent scientific and technical task. In doing so, the use of modern digital technologies within the framework of the industry 4.0 concept makes it possible to develop approaches that can significantly improve the quality of monitoring the state of the mine atmosphere at coal mining enterprises. This article provides a theoretical basis and test results for a system for continuous automatic monitoring of dust concentration in a mine atmosphere as the component of the multifunctional coal mine safety system. It is shown that monitoring the state of mine workings aerological safety can be carried out in real time through the system of the new generation using artificial intelligence. The ability of the proposed system to measure basic physical parameters affecting dust deposition (disperse composition, air humidity, dust concentration and air flow velocity) is noted.

**Keywords:** coal mine, continuous monitoring, mine atmosphere, aerological safety, multifunctional safety system, industry 4.0

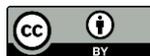

## 1. Introduction

В условиях роста угледобычи задача осуществления эффективного контроля запыленности и пылеотложения является важнейшей составляющей среди мероприятий, обеспечивающих промышленную и экологическую безопасность проведения горных работ в угольных шахтах.

С точки зрения промышленной безопасности взрывы газа и угольной пыли до сих пор остаются авариями с наиболее тяжкими последствиями в экономическом и социальном плане. Так за последние десятилетия на шахтах России, Индии и Китая практически во всех взрывах, имеющих крупномасштабные разрушения и приводящих к травмам и гибели работников предприятий, участвовала пылеметано-воздушная смесь.

В последние годы все большее давление на угольную промышленность оказывает экологическая политика. В частности, вопросы загрязнения окружающей среды твердыми частицами. Мельчайшие взвешенные частицы, образующие в результате производственной деятельности угольных предприятий, попадают в атмосферу близлежащих населенных пунктов, и без того сильно загрязненных промышленными выбросами и выхлопными газами. Ухудшение экологической обстановки неизбежно приводит к негативным последствиям: росту заболеваемости, в том числе онкологическими, ухудшению качества сельскохозяйственных земель, ухудшению видимости, снижение качества питьевой воды и другим.

Риск профзаболеваний пылевой этиологии еще остается высоким и социально значимым, что свидетельствует о наличии проблем в сфере условий, охраны труда и промышленной безопасности в отсутствие системного подхода [1]. Длительное воздействие пылевого аэрозоля приводит к профессиональным заболеваниям органов дыхания работников горных предприятий. При подземной добыче такие заболевания регистрируются у рабочих ведущих профессий, от 20,4 до 24,5 случая на 10 тыс., прошедших медицинский осмотр, и занимают второе место в структуре профессиональных заболеваний, помимо исторических интерстициальных заболеваний легких (пневмокониоз, силикоз и смешанный пылевой пневмокониоз). Шахтеры подвержены риску диффузного фиброза, связанного с пылью, и хронических заболеваний дыхательных путей, включая эмфизему и хронический бронхит.

Предотвращение взрывов пыли на угольных шахтах важно для обеспечения безопасности работников и предотвращения потенциальных чрезвычайных ситуаций. Для этого необходимо строго соблюдать правила безопасности, проводить регулярные инспекции и технические обследования оборудования, обучать персонал правилам обращения с пылью и применять специальные методы и технологии для ее контроля и снижения. Также необходимо обязательное использование защитных средств и эффективных систем вентиляции и очистки воздуха. Важно также проводить обучение и обучать работников эффективным методам контроля и предотвращения взрывов пыли в угольных шахтах.

Основной проблемой обеспечения безопасности является высокий уровень взрывоопасности из-за наличия угольной пыли. Угольная пыль может образовываться в результате различных процессов, таких как добыча, транспортировка и обработка угля. Она является серьезной угрозой для безопасности шахтеров из-за ее воспламеняемости и способности вызывать сильные взрывы. Поэтому контроль и предотвращение взрывов пыли является основной проблемой при обеспечении безопасности в угольных шахтах. Такие взрывы могут привести к гибели людей, разрушению оборудования и инфраструктуры, а также нанести значительный ущерб окружающей среде. Поэтому разработка и применение эффективных методов и технологий для контроля и предотвращения взрывов пыли является приоритетной задачей для безопасной работы угольных шахт.

Решение ее в разные периоды развития угольной отрасли формировалось в направлении создания комплексов профилактических мероприятий, которые регламентировались, на определенных этапах, различными нормативными документами.

## 2. Literature Review

Мониторинг угольной пыли и пылеотложений является актуальной задачей на сегодняшний день. Для повышения безопасности и исключения взрывов, горения угольной пыли [1] на шахтах важным аспектом является разработка объективного инструмента для продолжительного мониторинга, тем самым обеспечивая прогнозирование рисков взрыва пылеметановоздушной смеси и влияние угольной пыли на здоровье человека [2], [3].

В последние годы, с ростом использования техники, автоматизации и интеллектуальных систем в горнодобывающей промышленности, пылевое загрязнение становится более значительной проблемой, поэтому требуется более глубокое изучение вопросов мониторинга угольной пыли, осаждения пыли и предотвращения взрывов пылеметановоздушной смеси [4] с использованием современных технологий, основанных на искусственном интеллекте [5].

Известно, что самым распространённым способом борьбы с угольной пылью является вентиляция, хоть она и не относится к мониторингу пылеотложений но все же играет важную роль для безопасности горных выработок. Например, в исследовании [6] путем комбинирования метода численного моделирования было проанализировано влияние скорости потока на правила диффузии пыли источников пыли. Результаты исследования в свою очередь показали, что с увеличением подачи воздушных потоков массовая концентрация пыли снижается, но появляется риск, что данная скорость воздуха может вызвать повторный унос пыли, загрязняя рабочую среду. Также, в исследовании [7] для борьбы с угольной пылью было принято использовать технологию отвода принудительной вентиляции, где результаты применения показывают, что после принятия присоединенного воздуховода вдоль вентиляционного штрека формируется воздушная завесу, а после входа в секцию комбайна уже пылеподавляющая завеса. Когда соотношение осевого диаметра внешнего воздуха составляет 1:2-1:3, пыль можно контролироваться в пределах в пределах 17 мг/м$^3$, в свою очередь эффективность удаления пыли увеличивается при данной технологии до 93%. Но как было сказано раннее, вентиляция не является эффективным методом мониторинга пылеотложений и определения концентраций пыли, поэтому, для более эффективной борьбы с угольной пылью следует подобрать новые подходы для безопасной добычи полезного ископаемого.

Для формирования нового подхода к вышесказанной проблеме, потребовалось проанализировать имеющиеся материалы, в которых явно отражены проблемы осуществления эффективного контроля запыленности, пылеотложений и прогнозирования концентрации угольной пыли [8].

Тенденция развития заболеваний легких у горнорабочих, аварий на шахтах и запыленности атмосферы началась еще в прошлом веке. Так, например, в исследовании [9] рассматривается проблема вероятных причин существенного роста заболеваемости пневмокониозом. Целью данной работы являлось провести обзор респираторного осаждения, влияния угольной пыли на здоровье, мониторинга, нормативных требований и характеристик частиц. В данной работе был представлен обзор исследований, посвященных изучению вдыхаемой шахтной пылью, в которых основное внимание уделяется характеристике, респираторному осаждению и эффектам воздействия на здоровье. Кроме этого, при проведении исследования ав-тору потребовалось оценить точность систем мониторинга по осаждению и концентрации угольной пыли. Таким образом в работе не раскрыто влияние угольной пыли на здоровье шахтера.

В виду того, что каждое горное предприятие наращивает темпы добычи и стремится создавать безопасное производство, необходимо на ранних этапах внедрения систем просчитывать риски, обеспечивающие безопасность. Было проведено ряд исследований [10], в которых рассматривалось три модели классификации шахт, безопасности шахты и оптимальной вентиляции с помощью программных обеспечений MatLab и Lingo. Все эти исследования были связаны и осуществлены для того, чтобы дать надежные рекомендации для управления угледобывающими предприятиями высокого риска.

Для решения проблем сигнализации взрыва газа и угольной пыли и средств мониторинга в угольных шахтах, а также для повышения точности идентификации взрывов разработан метод на основе нейронных сетей [11]. Проанализированы коэффициенты звуков взрыва газа и угольной пыли, на основание чего создана модель извлечения признаков алгоритма Релифа. В результате эксперимента по распознаванию звука выяснилось, что разработанная модель алгоритмов способна точно различать каждый вид звука, а средняя скорость модели может достигать 95%, 95% и 95,8% соответственно и может соответствовать требованиям обнаружения взрывов угольных шахт и угольной пыли. Также было проведено исследование [12] в котором изучался вопрос борьбы с флегматизацией взрывов пылеметановоздушной смеси в шахтной атмосфере. Для достижения цели была разработана автоматизированная система управления и проведено наблюдения за ее работой в режиме реально времени. Благодаря разработанной автоматизированной системе было выявлено, что зависимости верхнего и нижнего порогов взрывоопасности пылеметановоздушной смеси и изучение воды и метана на молекулярной структуры и изучение воды и метана позволяют оценить эффективность мероприятий по предотвращению взрывов метана и пыли в угольных шахтах. Кроме того, для повышения точности необходимо повышение влажности чтобы система контроля флегматизации взрывов выдавала неискаженные данные.

В исследовании по непрерывному мониторингу угольной пыли [13] , проводились эксперименты в отдельно взятом забое о влиянии отрицательного воздействия угольной пыли на здоровье горнорабочих. Исследования проводились тремя методами с целью изучения и оценки концентраций угольной пыли. Результатом данного исследования было зафиксировано, что предельно допустимые концентрации в забое превышены в 10 раз и констатирована высокая концентрация вдыхаемой фракции угольной пыли, в результате чего эти фракции дольше всего остаются в организме и вызывают наибольшие негативные последствия для здоровья.

В дополнение к вышесказанному о влиянии вдыхаемых частиц угольной пыли на здоровье человека, были проведены экспериментальные испытания эффективности поверхностно-активных веществ для подавления угольной пыли [14]. Было установлено, что эффективность поверхностно-активного вещества существенно ухудшается при уменьшении крупности угольной пыли. Эффективность подавления угольной пыли размером от 0,1 до 1,0 мкм была лишь вдвое меньше, чем для пыли размером от 4 до 10 мкм. Основными факторами, влияющими на этот результат, являются шероховатость, удельная площадь поверхности, воздухопоглощаемость и количество частиц. Результатом исследования было то, что размеры частиц влияют на эффективность пылеподавления, а причиной всему этому то, что большая площадь поверхности, высокая способность к адсорбции воздуха и малое количеством частиц угля.

В свою очередь, мониторинг пылеотложений и концентрации угольной пыли возможно осуществлять с помощью ультразвукового контроля [15]. В данной статье предлагается разработка новой ультразвуковой измерительной системы, подходящей для надежного мониторинга изменений концентрации воздушной пыли. Система проходила испытания на опытно-промышленном классификаторе, предназначенном для воздушной сепарации и регенерации ультрадисперсных порошков из летучей золы сжигания угля . Метод мониторинга основан на измерении затухания ультразвуковых волн частицами, взвешенными в воздушном потоке классификатора. На примере шахты в Сербии, для мониторинга пылеотложений был предложен метод Гауссовской модели [16]. Данная модель проходила испытания на этапе открытия шахты и подтвердила обоснованность применения для будущего мониторинга, а зона, которая будет подвержена пылевому загрязнению, вызванному подготовительными или очистными работами, может быть предсказана гауссовской моделью.

Технология борьбы с угольной пылью на шахтах и рудниках стремительно продвигается вперед, так в [17] была разработана новая форсунка с вихревым сердечником основанная на механике жидкости и механизме коалесценции твердого тела и жидкости. Эффективная дальность распыления увеличивается с 5,2 до 5,9 м, а насыщенность распыляемого тумана значительно повышается. Тем самым, значительно повышается эффективность метода борьбы с пылью - следящего распыления, что важно для технологии предотвращения и контроля запыленности рудников с большой высотой выемки.

Также в работе [18] проводилось исследование для снижения концентрации пыли в шахтах. Для улучшения условий труда в качестве объекта исследования была задействована погрузочно-доставочная машина и установлена на ней система распыления. Исследование проводилось с помощью компьютерного моделирования и применяемыми уравнениями гидродинамики для моделирования полей скорости воздушных потоков и температуры воздуха. В результате исследования при помощи компьютерного моделирования было выявлено, что температура в рабочей зоне варьируется с 302-305 K, что является высоким показателем и негативно влияет на трудоспособность горнорабочих. Автор утверждает необходимость применения системы охлаждения для повышения теплового комфорта и снижения концентрации пыли опасной к взрыву. Кроме того, подобное исследование было выполнено на шахтах в Новом Южном Уэльсе, где на основе конкретных условий подземной выработки, выработанной комбайном, полученных с помощью пылемеров. построена трехмерная вычислительная модель, которая проверена на данных мониторинга пыли на месте [19]. Моделирование показало, что пыль высокой концентрации в основном распределяется по левой стороне комбайна, особенно на участке примерно в 3 м от забоя. Данные о воздушном потоке и миграции пыли из моделей были экспортированы для дальнейшей обработки данных с целью разработки иммерсивного учебного инструмента виртуальной реальности, позволяющего визуализировать данные о вентиляции и пыли, а также эффективно сообщать о мерах контроля пыли по охране труда и технике безопасности для комбайна непрерывного действия. Предлагаемая система является необходимой в целях безопасности ведения горных работ, а также является мощным инструментом для понимания передовых методов контроля пыли [20].

В исследовании [21] проведен обзор мониторинга твердых частиц угольной были на угольном разрезе Haerwusu, где исследовалась температура окружающей среды в декабре, январе и феврале и построены корреляционные зависимости характеристик вариации твердых частиц и их взаимосвязи с метеорологическими факторами. Из наблюдения авторов сказано, что концентрация твердых частиц наблюдается в декабре - январь - февраль. Температура положительно коррелирует с концентрацией твердых частиц в декабре, в то время как в январе она отрицательна. В то же время разница температур в декабре отрицательно коррелирует с концентрацией твердых частиц. При совместном действии нескольких метеорологических факторов величина воздействия на концентрацию ТЧ на дне карьера зимой определяется влажностью, температурой, скоростью ветра, разницей температур (обратная интенсивность температуры).

Мониторинг шахтной пыли является важным средством обеспечения безопасности производства и здоровья шахтеров. В статье [22] предлагается модель сервисного облачного мониторинга угольной пыли на основе облачной платформы. Благодаря данной системе работники предприятия смогут отслеживать состояние пыли в режиме реального времени и обеспечивать постоянный контроль. Благодаря долгосрочному применению системы накопления данных о профессиональных опасностях может быть использовано для общей базы данных, которая в будущем отразит совокупную дозу воздействия пыли на работников и вследствие чего снижения риска заболеваний на шахтах Китая.

В свою очередь в исследовании [23] было доказано, что измерительная система мониторинга с примененным бесконтактным методом пропорционального разностного обнаружения с двойным светом соответствует проектным требования, а точность и надежность системы доказаны экспериментально. Иной подход был предложен в [24] где было предложено использование технологии дистанционного зондирования для мониторинга заграгнения шахтной пылью. Исследование направлено на снижение рисков в шахте, безопасности ведения горных работ и здоровья обслуживающего персонала.

Также мониторинг угольной пыли возможно осуществлять с помощью оптической микроскопии [25]. Авторы статьи столкнулись с проблемой нехватки технологий мониторинга для подсчета и классификации угольной пыли. В связи с этим авторами был предложен новый подход, исследуя эффективность оптической световой микроскопии и поляризованного света. Результатом являлась демонстрация возможности использования световой микроскопии и поляризованного света с анализом изображения для подсчета и классификации типичных частиц пыли угольной шахты [26]. В конечном результате авторам не удалось дать точного ответа на исследование, так как мониторинг показал неоднозначные и многообещающие результаты, показывая, что в лабораторных условиях метод оптической микроскопии с автоматизированной обработкой изображений может обеспечить точное количественное определение угольных фракций.

Еще одним исследованием является использование недорогих датчиков для мониторинга угольной пыли [27]. В данной работе предлагается использовать недорогие датчики для мониторинга угольной пыли, так как в настоящее время технологии систем мониторинга являются дорогостоящими. За состоянием угольной пыли авторы предлагают применить светорассеивающие датчики твердых частиц, разработка которых за последние годы продемонстрировала определенные успехи. Для исследования использовались четыре вида датчиков Air trek, Gaslab, SPS30 и PMS5003. Результаты показали, что датчики Air trek и Gaslab оказались непригодными и показали плохую корреляцию. SPS30 перспективен только для низких концентраций (0-1,0 мг м⁻³), тогда как PMS5003 эффективно контролировал концентрации до 3,0 мг м⁻³. Данное исследование подтверждает о необходимости улучшения этих датчиков, тем самым облегчая их применение для улучшения безопасности и здоровья горняков.

Горные удары, усложнение горно-геологических условий разработки, взрывы метана и угольной пыли с увеличение глубины горных работ обуславливают необходимость создания эффективной многофункциональной системы безопасности, важнейшим фактором которого является аэрологическая безопасность [28].

В качестве решений для мониторинга и снижения риска взрывов при помощи многофункциональной системы безопасности является метод оперативного определения уровня осажденной пыли в горных выработках, позволяющий предотвратить ее возможный взрыв [29]. В исследовании определен интервал проведения мероприятий по обеспечению безопасности горных выработок для предотвращения рисков взрыва пылеметановоздушной смеси при забойных работах высокопроизводительным оборудованием. Однако, эффективное внедрение новых технологических процессов невозможно без точного и оперативного контроля уровня пылеобразования в горных выработках, особенно в тех случаях, когда на забой возрастает воздействие, так как необходимо проводить дополнительный анализ фракционного состава угольной пыли во избежание взрывов и самовозгорания метанопылевой смеси [30], [31].

Для точного анализа и прогноза концентрации шахтной пыли проведено исследование, которое основано на модели Грея Маркова [32]. Модель Грея Маркова была применена для прогнозирования концентрации шахтной пыли и сравнена с результатами прогнозирования модели нейронной сети, модели серого прогнозирования и модели ARIMA. Результаты показали, что модель Грея Маркова показала более точные результаты по сравнению с другими моделями. Таким образом, была проверена точность и рациональность данной модели для прогнозирования концентрации шахтной пыли.Для того чтобы избежать катострафических последствий в шахте, в работе [33] проведено исследование контроля продолжительности пламени взрыва и интенсивности сигнала пламени, что позволило разумно определить количество и длительность впрыскиваемого ингибитора, являющимся ключом к сдерживанию распространения взрыва. Результаты потребовались для выяснения механизма распространения пламени реакции взрыва пылеотложений при взрыве газа и предотвращения газопылевых взрывных аварий.

Другой проблемой, вследствие чего образуется пожар являются выбросы окиси углерода CO. Для мониторинга подземных пожаров существует статистический контроль процессов шахтных пожаров. Путем анализа изменения содержания окиси углерода в зависимости от времени в подземных шахтах с использованием этого метода была создана эффективная методология борьбы с угольными пожарами [34].

К традиционным методам контроля пылеобразования и пылеотложений относятся радиоизотопные, оптические и химические методы. Данные методы являются дорогостоящими и кроме того, погрешность измерений составляет 10%. Поэтому в исследовании [35] был исследован и разработан новый метод лишенный недостатков приведенных выше. Осуществлять контроль за пылеобразованием предлагается новым методом – термогравиметрическим. Также, были проведены экспериментальные исследования. Выделены непересекающиеся интервалы термогравиметрической реакции: выход влаги (35-132 °С); выход летучих веществ из угля (380-580 °С); термическая деструкция известняка с выходом углекислого газа (650-850 °С). На основании экспериментальных данных были рассмотрены математические зависимости для обработки характеристик термогравиметрических кривых при определении содержания негорючих компонентов в пробе шахтной пыли.

Для сравнения методов мониторинга было проведено исследование на закрытой шахте в Канаде [36], где в качестве мониторинга выступали спутниковые снимки, анализ тканей лишайника и пассивные коллекторы сухих отложений с думя различными конфигурациями, датчики осаждения пыли. Было показано, что самыми эффективными инструментами для определения степени загрязненности являлись *лишайники и спутниковые снимки*. Также отмечено, что самое высокое осаждение пыли было выявлено в зимние месяцы, а самые низкие в летние месяцы, это связано с тем, что в в зимних погодных условиях наиболее усиленная эрозия.

Однако, чтобы повысить уровень мониторинга, в исследовании [37] разрабатывается система на основе интернета вещей IoT. В работе с помощью контроллера предлагается концепция интернета беспроводной сенсорной сети, которая поможет отслеживать температуру, влажность и газ в шахте. Имеющиеся системы наблюдения представляют собой регулярно подключенные системы, которые играют ключевую роль в безопасности угольных шахт, а с непрерывным расширение зон добычи данные системы с трудом справляются с поставленной задачей обеспечения безопасности. Кроме того, прокладка кабелей является дорогостоящей и трудоемкой. Поэтому, создается система мониторинга безопасности угольных шахт на основе беспроводной сенсорной сети и интернета вещей, которая повысит безопасность, уменьшит риск взрывов и будет осуществлен контроль пылеотложений. В аналогичном исследовании [38] также на основе IoT предлагается комплексный подход чтобы повышают уровень интеллекта при удалении пыли из угольных шахт. В исследовании описываются ключевые аспекты и этапы создания базы данных основных элементов для удаления пыли. Кроме того, в нем подчеркивается практическое применение методов, которые необходимы для решения задач удаления угольной пыли.

В качестве примера будет уместно затронуть мониторинг угольный пыли и пылеотложения карьера. По сравнению с подземной добычей, открытая добыча имеет множество источников пыли, а зона распространения может привести к большим колебаниям концентрации пыли. Таким образом, в статье [39] предлагается гибридная модель для прогнозирования концентрации угольной пыли. *Для оценки и точности модели потребовалось сравнить три модели ARIMA, LSTM и модель LSTM-Attention* в абсолютно идентичных условиях. Результат показал, что модель LSTM-Attention более стабильна и имеет точность прогнозирования выше приведенных моделей. В исследовании отмечено, что данную модель можно применять в условиях открытых горных работ для прогнозирования концентрации пыли.

Вышеупомянутые проблемы с пылеотложением и высокой концентрацией угольной пыли, ее образованием и распространением в угольных шахтах и горных выработках были решены в научных трудах [40] [41]. Решение проблемы с угольной пылью было представлено в виде разработанной системы непрерывного автоматического контроля запыленности и интенсивности пылеотложений - прибора СКИП, который в режиме реального времени с помощью системы нового поколения с использованием искусственного интеллекта позволяет проводить измерения базовых физических параметров, влияющих на пылеотложение. Прибор является новейшей разработкой ученых Российской Федерации который прошел лабораторные и шахтные испытания и подтвердил правильность разработанной физической модели.

## 3. Theoretical basis for controlling dust deposition in mine workings

### 3.1. General Information

### 3.2. System Configurations Assessment

**4. Results**

**5. Discussion**

**6. Conclusions**